\documentclass{PoS}

\usepackage{graphicx,color}
\usepackage{amsmath,amssymb,bm}

\def\XXint#1#2#3{{\setbox0=\hbox{$#1{#2#3}{\int}$}
     \vcenter{\hbox{$#2#3$}}\kern-.5\wd0}}

\usepackage{amsmath}
\usepackage{graphicx}
\usepackage{mathptmx}      
\usepackage{latexsym}
\usepackage{bm}
\usepackage{comment}

\usepackage{supertabular} 
\usepackage{placeins}
\usepackage{epsfig}
\usepackage{graphicx}

\usepackage{epstopdf}

\graphicspath{{/home/pgortega/Dropbox-ort/Documentos_Investigacion/Articulos/XCount/}}

\title{Does $X(3872)$ count?}

\ShortTitle{Does $X(3872)$ count?}

\author{\speaker{Pablo G. Ortega} \thanks{This work has been partially funded by the Spanish Ministerio de
Economia y Competitividad and European FEDER funds (Grant
No. FIS2014-59386-P, FPA2016-77177-C2- 2-P), the Agencia de Innovacion
y Desarrollo de Andalucia (Grant No.  FQM225), and by Junta de
Castilla y Le\'on and European Regional Development Funds (ERDF) under
Contract no. SA041U16. 
}\\
        Grupo de F\'isica Nuclear and Instituto Universitario de F\'isica 
Fundamental y Matem\'aticas (IUFFyM), Universidad de Salamanca, E-37008 
Salamanca, Spain\\
        E-mail: \email{pgortega@usal.es}}

\author{Enrique Ruiz Arriola\\
        Departamento de
  F\'{\i}sica At\'omica, Molecular y Nuclear  and Instituto Carlos I
  de F{\'\i}sica Te\'orica y Computacional \\ Universidad de Granada,
  E-18071 Granada, Spain.\\
        E-mail: \email{earriola@ugr.es}}

\abstract{ The question on whether or not weakly bound states should
  be effectively incorporated in a hadronic representation of the QCD
  partition function is addressed by analyzing the example of the
  $X(3872)$, a resonance close to the $D\bar D^*$ threshold which has
  been suggested as an example of a loosely bound molecule. This can
  be decided by studying the $D \bar D^*$ scattering phase-shifts in
  the $J^{PC}=1^{++}$ channel and their contribution to the level
  density in the continuum, which also gives information on its
  abundance in a hot medium. In this work, it is shown that, in a
  purely molecular picture, the bound state contribution cancels the
  continuum, resulting in a null occupation number density at finite
  temperature, which implies the $X(3872)$ does not count below the
  Quark-Gluon Plasma crossover ($T \sim 150$MeV).  However, if a
  non-zero $c \bar c$ component is present in the $X(3872)$ wave
  function such cancellation does not occur for temperatures above
  $T\gtrsim 250$MeV.}

\FullConference{XVII International Conference on Hadron Spectroscopy and Structure - Hadron2017\\
		25-29 September, 2017\\
		University of Salamanca, Salamanca, Spain}

\begin{document}

QCD thermodynamics in a finite box is closely related to
counting hadronic states.  Within Hadron Resonance Gas (HRG)
approximation, QCD thermodynamics in the confined phase is reduced due
to the role of narrow resonances~\cite{Dashen:1974jw} and effective
elementarity~\cite{Dashen:1974yy}.  Also, if we build the cumulative
number of all the states listed in the Particle Data Group (PDG),
$N_{\rm PDG}(M)$, we can check that it fits the trace anomaly,
$\epsilon- 3 P = T^5 \partial_T (\log Z /T^3)/V$, on the lattice at
temperatures $T \lesssim 200 {\rm MeV}$ below the crossover to the
Quark-Gluon Plasma (QGP) phase~\cite{Borsanyi:2016ksw} (see also
Ref.~\cite{Arriola:2014bfa} and references therein).

These results point to the need to count all the states listed in the
PDG as genuine contributions to the QCD partition function and, thus,
also included in the HRG.  However, under special circumstances such
strong conclusion is incorrect, as loosely bound states may become
fluctuations in a mass-spectrum coarse grained
sense~\cite{Dashen:1974ns}.  These authors concluded that certain
interactions do not create new states but just reorder the already
existing ones.  For the deuteron, a $J^{PC}=1^{++}$ np composite, the
nearby np continuum compensates the weak binding effect and the
overall contribution of the deuteron is zero as depicted in
\cite{Arriola:2015gra}.

The recent discovery of the so-called X,Y,Z states arises the question
if these states must be included in the PDG and whether or not they
add redundancy when building the hadron
spectrum~\cite{Arriola:2014bfa,Arriola:2015gra}.  In this work we
study this effect for the $X(3872)$ resonance, a $1^{++}$ state
discovered in 2003 by the Belle Collaboration~\cite{Choi:2003ue},
whose properties point to a dominant $D\bar D^*$-molecular structure.
Analyzing $D \bar D^*$ scattering we will show that the answer to this
question depends on the particular dynamics of the
system~\cite{Ortega:2017hpw}.

According to the quantum virial expansion~\cite{Dashen:1969ep}, the
average density of a composite particle with $g$-degrees of freedom
and mass $m$ in a medium with temperature $T$ can be written as
\begin{eqnarray}
n(T) = \int \frac{d^3 p}{(2\pi)^3} dm \frac{g}{e^{\sqrt{p^2+m^2} /T} + \eta} \rho(m)\,\,\mbox{, where }\,\,\rho(m)=\frac1{\pi} \frac{d \delta}{d m},
\end{eqnarray}
and with $\delta$ the scattering phase shifts. This equation
implicitly includes the contribution of elementary particles, since
the phase-shift of a narrow resonance with mass $m_R$ and width
$\Gamma_R \to 0$ can be written as $\delta(m)=
\tan^{-1}[(m-m_R)/\Gamma_R]$, so that $\delta'(m) \to \pi
\delta(m-m_R)$~\cite{Dashen:1974jw}.

Now, for a certain type of weakly bound states their contribution may
in fact vanish, as was shown by Dashen and
Kane~\cite{Dashen:1974ns}. The cumulative number in a given channel in
the continuum with threshold $M_{\rm th}$ is
\begin{eqnarray}
N(M)= \sum_n \theta(M-M_n^B)
  + \frac1\pi \sum_{\alpha=1}^{K} [\delta_\alpha (M)-\delta_\alpha (M_{\rm th})] \,,
\label{eq:ncum} 
\end{eqnarray}
where we have explicitly separated the bound state $M_n^B$
contributions from the continuum, written in terms of the eigen
phase-shifts of the coupled channel S-matrix. This equation satisfies
$N(0)=0$ and, in the single channel case, when $M \to\infty$ it
becomes $N(\infty)=n_B + [\delta(\infty)-\delta(M_{\rm th})]/\pi=0.$
This is a consequence of Levinson's theorem, which states that the
total number of states does not depend on the interaction.

Since its discovery, the weak binding of the $X(3872)$ has suggested a
purely molecular nature with no reference to underlying
quarkdynamics (see e.g. \cite{Gamermann:2009uq}). Within
this molecular picture, in the $D\bar D^*$ channel the appearance of
the $X(3872)$ rapidly shifts $M= M_{\rm th} -B_X $ by one unit so that
$N ( M_{\rm th}-B_X +0^+)- N ( M_{\rm th}-B_X -0^+)=1 $. However, this
number decreases slowly to zero at about $\Delta M_{D \bar D^*} \sim
200$ MeV, so $N ( M_{\rm th}+ \Delta M_{D \bar D^*})- N ( M_{\rm
  th}-B_X -0^+) \sim 0 $~\cite{Ortega:2017hpw}.  This illustrates the
point made by Dashen and Kane~\cite{Dashen:1974ns}, showing that, in
the purely molecular picture, the $X(3872)$ does not count in the $D
\bar D^*$ continuum on coarse mass scales of about $200$ MeV.

This situation may change if the inner structure of the $X(3872)$
includes a non-vanishing $c\bar c$ component.  The multichannel
scattering problem with confined intermediate states was initiated
after the first charmonium
evidences~\cite{Dashen:1976cf,Eichten:1978tg} based on the
decomposition of the Hilbert space as ${\cal H} = {\cal H}_{c \bar c}
\oplus {\cal H}_{D \bar D}$. Such decomposition was implemented in
Ref.~\cite{Ortega:2009hj,Ortega:2012rs} in the framework of a
widely-used constituent quark model (CQM)~\cite{Vijande:2004he}.
There, a coupled-channels calculation for the $J^{PC}=1^{++}$
states was addressed, and the $X(3872)$ was
described as a mostly $D\bar D^\ast+h.c.$ molecule with a sizable
amount of $c\bar c(2^3P_1)$ state, while an additional resonance was
found with more than $60\%$ of $c\bar c$ structure, assigned to the
$X(3940)$.  The meson-meson interaction includes the exchange of
pseudo-Goldstone bosons at $q\bar q$ level~\cite{Vijande:2004he} and
the coupling with two- and four-quark configurations through the
$^{3}P_{0}$ model.  From the latter transition mechanism, an effective
potential $V^{\rm eff}_{\beta'\beta}$ arises, encoding the coupling
with the $c\bar c$ bare spectrum (see further details in
Ref.~\cite{Ortega:2012rs}).  The intensity of the $^3P_0$-model is
controled by a dimensionless parameter, dubbed $\gamma$, originally
constrained via strong decays in the charmonium spectrum.
Here, the effect of adiabatically connect the $c\bar c$ spectrum and the $D\bar D^\ast$ channel is analyzed,
thus the $\gamma$ will be varied from zero to the value employed in Ref.~\cite{Ortega:2009hj}, fixing the mass of the 
bound state $X(3872)$ at its experimental value of $3871.7$ MeV. 

\begin{figure*}[b]
\begin{center}
\includegraphics[width=0.45\linewidth]{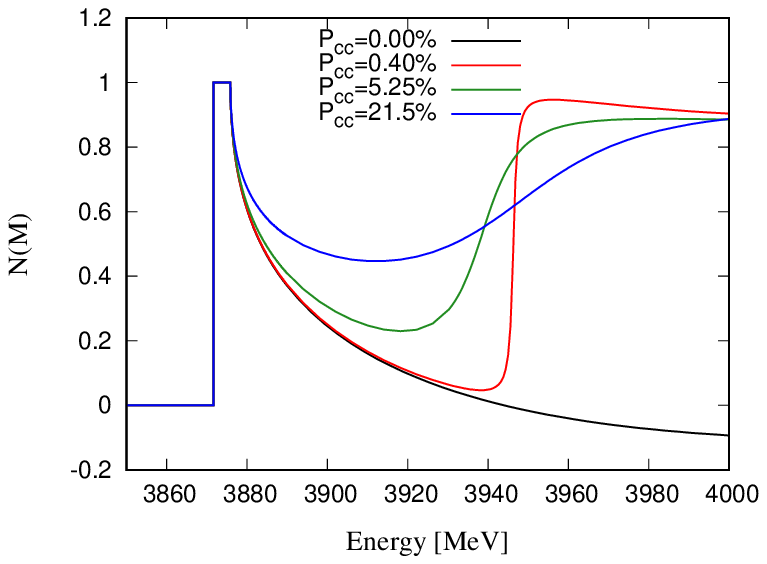}
\includegraphics[width=0.45\linewidth]{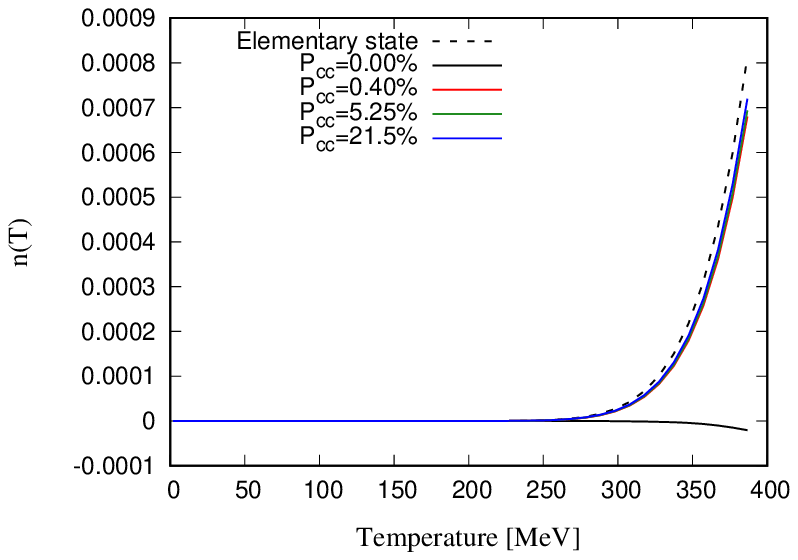}
\end{center}
\caption{Left panel: Cumulative number in the $X(3872)$ channel as a
  function of the $D \bar D^*$ mass. Right panel: Occupation number $n(T)$ of the $D
  \bar D^*$ in the $J^{PC}=1^{++}$ channel as a function of the
  temperature T (in MeV). The dashed line represents the contribution
  of the $X(3872)$ assuming it is an elementary particle and no
  continuum contribution.}
  \label{fig:ncluster}
\end{figure*}

The cumulative number of states is shown in
Fig.~\ref{fig:ncluster}(a).  We check that the conclusion of Dashen
and Kane~\cite{Dashen:1974ns} is maintained only when the $X(3872)$
meson is a pure molecule, so in this case the $X(3872)$ does not
count.  On the contrary, there is an outstanding turnover of the
cumulative number as soon as we couple with the $c\bar c$ structures,
even for small couplings. Such steep rise in the phase shifts points
to a resonance located at a mass of $M\sim3945$MeV, which was
identified with the $X(3940)$ resonance in Ref.~\cite{Ortega:2012rs}.
In the purely molecular picture the $c\bar c$ structures decouple from
the $D\bar D^\ast$ states and such resonance disappears.  So, we can
conclude that such raise is not a consequence of the $X(3872)$ but it
is due to the onset of the $X(3940)$ resonance.

Finally, we discuss the implications for finite temperature
calculations.  The occupation number is shown in
Fig.~\ref{fig:ncluster}(b), where we can appreciate the cancellation
between the continuum and the bound state only happens for zero $c\bar
c$ content, when both two- and four-quark sectors are decoupled.
However, we want to remark that the non-vanishing ocupation number is
basically due to the resonant reaction $D \bar D^* \to X(3940) \to D
\bar D^* $.

Concluding, our study shows that the signal for $X(3872)$ abundance may in fact be erroneously confused
with the $X(3940)$ as a non-vanishing occupation number of the $D \bar
D^*$ spectrum in the $1^{++}$ channel at temperatures, $T \gtrsim 250$
MeV, above the crossover to the QGP phase. Below this temperature, the
$X(3872)$ does not count and should not be included in the Hadron
Resonance Gas.

\bibliographystyle{JHEP}

\bibliography{X3872}

\end{document}